\documentclass{ws-ijmpa}

\begin{document}

\markboth{B.A. Arbuzov and I.V. Zaitsev}
{Consistent interpretation of diboson excesses}

%
%

\title{On a possibility of a consistent interpretation\\ 
of diboson excesses at LHC}

\author{Boris A. Arbuzov}

\address{\it M.V. Lomonosov Moscow State University,\\
Leninskie gory 1, 119991 Moscow, Russia\\
arbuzov@theory.sinp.msu.ru}

\author{Ivan V. Zaitsev}

\address{\it M.V. Lomonosov Moscow State University,
\\Leninskie gory 1, 119991 Moscow, Russia\\
zaitsev@theory.sinp.msu.ru}

\maketitle

\begin{history}
\received{Day Month Year}
\revised{Day Month Year}
\end{history}

\begin{abstract}
Recently reported diboson and diphoton excesses at LHC are interpreted  to be connected with heavy $WW$ zero spin resonances. The resonances appears
due to the wouldbe anomalous triple interaction of the weak bosons, which is defined by well-known coupling constant $\lambda$. The $2\, TeV$ anomaly tentatively corresponds to weak isotopic spin 2 scalar state and the
$\gamma\gamma\,\,750\, GeV$ anomaly corresponds to weak isotopic spin 0
pseudoscalar state.
We obtain estimates for the effect, which qualitatively agree with ATLAS data. Effects are predicted in
a production of $W^+ W^-, (Z,\gamma) (Z,\gamma)$ via
resonance $X_{PS}$ with $M_{PS} \simeq 750\,GeV$, which
could be reliably checked at the upgraded LHC at $\sqrt{s}\,=\,13\, TeV$.
In the framework of an approach to the spontaneous generation of the triple anomalous interaction its coupling constant is estimated to be  $\lambda = -\,0.02\pm 0.005$ in an agreement with existing restrictions. Specific
predictions of the hypothesis are significant effects in decay channels
$X_{PS} \to \gamma\,l^+\,l^-\,, X_{PS} \to l^+\,l^-\,\,l^+\,l^-\,(l = e,\,\mu)$.


\end{abstract}

\ccode{PACS numbers: 12.15.Ji; 12.60Cn; 14.70.Fm; 14.70.Hp}
\section{Diboson anomalies}
In experiments~\cite{ATLAS} indications for excesses in the production of boson pairs  $WW,\,WZ,\,ZZ$ were observed
at invariant mass $M_R\simeq 2\,TeV$. Data for these processes are also present in works~\cite{CMSE1,CMSE2}. Despite the fact that the wouldbe effect is not finally established yet, the publication causes numerous
proposals for an interpretation mostly in terms of theories beyond the Standard Model (see, {\it e.g.}~\cite{INTER7}). 
There are most recent data on $\gamma\,\gamma$ anomaly at
$M(\gamma\,\gamma) \simeq 750\,GeV$~\cite{750A,750C}, which also caused
numerous proposals for an interpretation~\cite{ellis}. As a rule for 
interpretations of the effects the {\it ad hoc} proposals are expressed, each  being based on a variant of the Physics beyond the Standard Model. 

However, it seems to be quite advisable to look for options being less radical, which are closely connected with our current knowledge. In this way we 
are to consider a possibility to explain the anomalies in the framework of 
non-perturbative contributions in the Standard Model.
Then for the heavy resonance~\cite{ATLAS} the idea naturally appears to prescribe the phenomenon to a resonant state of two $W$-s~\cite{AZ15}. 

Of course
the standard perturbative approach fails for application to such states. However the same assertion is true for usual hadronic states, {\it e.g.}
light mesons,  
and here there are non-perturbative approaches, which allows to
achieve successes in such problems. As for the light meson physics, we may
refer to the well-known non-perturbative effective Nambu -- Jona-Lasinio interaction as a tool for an
adequate description of the problem. $W\,W$ bound states and resonances also could appear provided an effective interaction of $W$-s does exist. Such
three-boson anomalous effective interaction is not an {\it ad hoc} hypothesis. It was proposed a long time ago~\cite{Hag1,Hag2} and was extensively studied
experimentally. 

We have studied this option in the framework of a theoretical 
approach to a possibility of a spontaneous generation of effective interactions~\cite{BAA09,AVZ2}. It is important, that the intensity of the interaction grows with the energy scale upgrowth. The interaction becomes 
strong enough just on the scale of few TeV, so the appearing of heavy $W\,W$
resonance becomes natural~\cite{AZ15}.
Namely, in work~\cite{AZ15} we have considered interpretation of the effect in terms
of a weak isotopic spin 2 scalar $W\,W$ state. Indeed, pair of triplets $W^a$ could form
a resonance state, the so-called $W$-ball. Of course the well known gauge interaction of these
bosons with coupling $g(M_W)=0.65$ can not bind them in the resonance state
with mass being of a $TeV$ scale. However, as we already mentioned, provided there exists also the  additional effective interaction~\cite{Hag1,Hag2}, we 
come to a formation of a resonance under discussion.

Could this approach also be applied to
analogous states? We would consider this problem in the present paper.

\section{A model for the $WW$ resonance}

Now let us consider a possibility of a heavy pseudoscalar resonance
in case of an existence of the anomalous three-boson interaction, which in
conventional notations~\cite{Hag1,Hag2} looks like
\begin{eqnarray}
& &-\,\frac{G}{3!}\,F\,\epsilon_{abc}\,W_{\mu\nu}^a\,
W_{\nu\rho}^b\,
W_{\rho\mu}^c\,;\quad G\,=\,-\,\frac{g\,\lambda}{\,M_W^2}
\label{FFF}\\
& &W_{\mu\nu}^a\,=\,
\partial_\mu W_\nu^a - \partial_\nu W_\mu^a\,+g\,\epsilon_{abc}
W_\mu^b W_\nu^c\,;\nonumber
\end{eqnarray}
where $g \simeq 0.65$ is the electro-weak coupling.
The best limitations for parameter $\lambda$ read~\cite{PDG}
\begin{equation}
\lambda_\gamma = -\,0.022\pm0.019\,;\quad
 \lambda_Z = -\,0.09\pm0.06\,; \label{lambda1}
\end{equation}
where a subscript denote a neutral boson being involved in the experimental definition of $\lambda$.
Let us emphasize that
$F \equiv F(p_i)$ in definition~(\ref{FFF}) denotes a form-factor, which is either postulated as
in original works~\cite{Hag1,Hag2} or it is just uniquely defined as
in works on
a spontaneous generation of effective interaction~(\ref{FFF})
~\cite{BAA09,AVZ2}. In any case the form-factor guarantees the effective interaction to act in a limited region of the momentum space. That is
it vanishes for momenta exceeding scale $\Lambda_0$. Formfactor $F$ is
explicitly shown {\it e.g.} in work~\cite{AZ15}.
Calculations were done in the framework of an approximate scheme, which accuracy was estimated to be $\simeq (10 - 15)\%$~\cite{BAA04}. Would-be existence of effective interaction~(\ref{FFF}) leads to important non-perturbative effects in the electro-weak interaction.

In particular, one might expect resonances to appear in the system of
two $W^a$-bosons. A possibility of an appearance of such states (W-balls)
was already discussed, {\it e.g.} in works~\cite{AVZ2,AZ12}.
In the previous work~\cite{AZ15} we have studied the $2000\, GeV$ anomaly and
have came to the conclusion, that data may be described in terms of a weak
isospin 2 scalar resonance. The effect is due to anomalous interaction~(\ref{FFF}) and we come to a conclusion~\cite{AZ15}, that there is a possibility to describe data~\cite{ATLAS} with
\begin{equation}
\lambda\,=\,-\,0.017 \pm 0.005.\label{eq:lam1}
\end{equation}

Let us turn to recent indications for existence of the other effect: the
$\gamma \gamma$ enhancement at invariant mass $M_{\gamma\gamma} \simeq 750\,GeV$.
We would consider this effect being explained by existence of zero weak isotopic spin pseudoscalar state $X_{PS}$, which interaction with electroweak bosons is described by the following effective expression
\begin{equation}
L_{eff}\, = \,\frac{G_{PS}}{4}\, \delta_{ab}\,\epsilon^{\,\mu \nu \rho \sigma}
W^a_{\mu\nu}\, W^b_{\rho\sigma}\,X_{PS};
\label{GXWW}
\end{equation}
Let us  consider a Bethe-Salpeter equation for a pseudoscalar resonance consisting of two $W$ corresponding to the weak isospin: $I = 0$.
\begin{figure*}
\begin{center}
\includegraphics[scale=0.4]{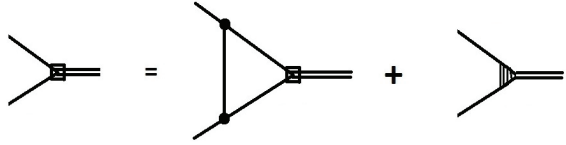}
\caption{Diagram form of equation~(\ref{eq:BS2}). Simple lines represent
$W$-s, a double line corresponds to the resonance, black circles correspond to
interaction~(\ref{FFF}), squares correspond to BS wave function. The triangle corresponds to vertex~(\ref{GXWW}).}
\label{fig:compenG}
\end{center}
\end{figure*}
With interaction~(\ref{FFF}) we have the following Bethe-Salpeter
equation for state $X_{PS}$ in correspondence to diagrams
presented in Fig 1 under assumption of existence of interaction~(\ref{GXWW}).
\begin{eqnarray}
& & \Psi_{PS} =G_{PS} + \frac{G^2}{8\,\pi^2}\biggl(
\frac{1}{6 x}\int_0^x \Psi_{PS}(y) y^2 dy -\frac{1}{2}\int_0^x \Psi_{PS}(y) y dy-\nonumber\\
& &\frac{x}{2}\int_x^{\infty} \Psi_{PS}(y) dy + \frac{x^2}{6}\int_x^{\infty} \frac{\Psi_{PS}(y)} {y} dy\biggr);\label{eq:BS2}
\end{eqnarray}
where coupling constant $G_{PS}$ is defined by~(\ref{GXWW}).
Here in view of large value $M_X \simeq 0.75\,TeV$ of the wouldbe resonance we neglect $W$ mass. With the following definition of variables
\begin{equation}
z = \frac{G^2\,x^2}{128\,\pi^2};\; t = \frac{G^2\,y^2}{128\,\pi^2};\; \label{define}\\
\end{equation}
we come to the following equation
\begin{eqnarray}
& & \Psi_{PS}(z) = G_{PS} +
\frac{4}{3 \sqrt{z}}\int_0^z \Psi_R(t) \sqrt{t} dt - 4 \int_0^z \Psi_{PS}(t) dt-\nonumber\\
& &4 \sqrt{z}\int_z^{\infty} \frac{\Psi_{PS}(t)}{\sqrt{t}} dt +\frac{4 z}{3}
\int_z^{\infty} \frac{\Psi_{PS}(t)} {t} dt.\label{eq:BSZ}
\end{eqnarray}
Equation~(\ref{eq:BSZ}) satisfies condition
\begin{equation}
\Psi_{PS}(0)\,=\,G_{PS}.\label{eq:norm}
\end{equation}

By successive differentiations of equation~(\ref{eq:BSZ}) we obtain a Meijer
differential equation for function $\Psi_{PS}(z)$
\begin{eqnarray}
& &\biggl[\biggl(z\frac{d}{dz}+\frac{1}{2}\biggr)\biggl(z\frac{d}{dz}\biggr)
\biggl(z\frac{d}{dz}-\frac{1}{2}\biggr)\biggl(z\frac{d}{dz}-1\biggr)+
z\biggr]\,\Psi_{PS}(z)\,=\,0.\label{eq:diff}
\end{eqnarray}
Then the solution, which fulfill boundary condition both at zero and at the infinity is expressed in terms of Meijer functions (see {\it e.g.}~\cite{be,ABOOK}) in the following way
\begin{equation}
\Psi_{PS}(z)\,=\,\frac{G_{PS}}{2}\,G_{04}^{30}\bigl( \,z\,|_{\;0,1/2,1,-1/2}\bigr).\label{eq:solC}
\end{equation}
\begin{figure*}
\begin{center}
\includegraphics[scale=0.6]{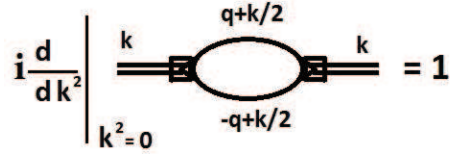}
\caption{Diagram representation of normalization condition for
coupling $G_{PS}$. Simple lines represent
$W$-s, a double line corresponds to the resonance, squares correspond to BS wave function.}
\label{fig:norm}
\end{center}
\end{figure*}

The normalization condition for Bethe-Salpeter wave function~(\ref{eq:solC}) give according to diagram Fig.~\ref{fig:norm} with account of definition~(\ref{define}) the following relation
\begin{eqnarray}
& &\frac{9}{64\,\pi^2}\int_0^{\infty} dy\, \Psi_{PS}(y)^2\,=\,
\frac{9\,\sqrt{2}\,G_{PS}^2}{16 \pi\, G}\,I\,=\,1;\label{eq:norm}\\
& &I\,=\,\int_0^{\infty}\frac{G_{04}^{30}\bigl( \,t\,|_{\;0,1/2,1,-1/2}\bigr)^2}{2 \sqrt{t}}dt\,
\,=\,\frac{\pi}{8}.\nonumber
\end{eqnarray}
With values $I$~(\ref{eq:norm}), $g = 0.65$ and with account of~(\ref{FFF}) we obtain coupling $G_{PS}$
\begin{equation}
G_{PS}\,=\,\frac{8}{3\,M_W}\sqrt{ g\,|\lambda|\,\sqrt{2}}\,=\,0.00389\,
\frac{1}{GeV};\label{GX}
\end{equation}
where numerical value corresponds to $\lambda = \lambda_0 = -0.015$, that is safely
inside restrictions~(\ref{lambda1}) and estimates~(\ref{eq:lam1}). Value~(\ref{GX}) corresponds to scale $\Lambda \simeq 0$. We take
value~(\ref{GX}) for estimates of effects, bearing in mind, that
for other values of scale $\Lambda$ coupling $G_{PS}$ is defined by 
solution~(\ref{eq:solC}), namely
\begin{equation}
G_{PS}(\Lambda)\,=\,\frac{G_{PS}}{2}\,G_{04}^{30}\bigl( \,z_{\Lambda}\,|_{\;0,1/2,1,-1/2}\bigr);\quad z_\Lambda = \frac{G^2\,\Lambda^4}{128\,\pi^2}.\label{eq:GG}
\end{equation}
We evaluate the pseudoscalar resonance decay probabilities with $\Lambda_D = M_{PS} = 750\,GeV$. 
For estimations of cross sections we take $\Lambda$ in correspondence to 
maxima of structure functions. That is $\Lambda(\sqrt{s}) \simeq \sqrt{s}/7$. Then for the decay and for two values $\sqrt{s} = 8\,TeV$ and $\sqrt{s} = 13\,TeV$ we have the following values for effective coupling $G_{PS}$
\begin{eqnarray}
& & G_{PS}(\Lambda_D) = G_{PS}(0.75\,TeV) = 0.00362;\nonumber\\
& &G_{PS}(8\,TeV/7)\,=\,0.00332;\quad  G_{PS}(13\,TeV/7)\,=\,0.00262.\label{eq:8;13}
\end{eqnarray}

Thus we have interaction~(\ref{GXWW}) with parameters $G_{PS}$~(\ref{GX},\ref{eq:8;13}) and $M_{PS} = 750\,GeV$. We use well-known relation
\begin{equation}
W^0 = \cos\theta_W\,Z\,+\,\sin\theta_W\,A;\label{Weinberg}
\end{equation}
and obtain for partial decay widths of the pseudoscalar $X_{PS}$
\begin{eqnarray}
& &\Gamma(W^+\,W^-)\,=\,51.3\,GeV\;(42.9\%);\quad \Gamma( Z\,Z)\,=\,14.8\,GeV\;(12.4\%);\nonumber\\
& &\Gamma(Z\,\gamma)\,=\,7.8\,GeV\;(7.6\%);\quad
\Gamma(\gamma\,\gamma)\,=\,1.5\,GeV\;(1.2\%);\label{WIDTH}\\
& &\Gamma(W^+\,W^-\,Z)\,=\,35.2\,GeV\;(29.5\%);\quad \Gamma( W^+\,W^-\,\gamma)\,=\,7.3\,GeV\;(6.1\%);\nonumber\\
& & \Gamma_t(X_{PS})\,=\,119.4\,GeV.\nonumber
\end{eqnarray}
We would present also probabilities for the following specific channels,
where $l$ means light lepton ($\mu,\,e$)
\begin{equation}
\Gamma(l^+\,l^-\gamma)  = 0.67\,GeV\,
(0.64\%);\;\Gamma(l^+\,l^-\,l^+\,l^-)= 0.068\,GeV (0.065\%).\label{ll}
\end{equation}
Then we calculate cross sections for $X_{PS}$ production in $p\;p$
collisions for $\sqrt{s} = 8\,TeV$ and for $\sqrt{s} = 13\,TeV$ using values~(\ref{eq:8;13}). In doing
this we use the CompHEP package~\cite{CompHEP}.

\begin{table}
\begin{center}
Table 1\\

Results for cross-sections of  $X_{PS}$ production in $p+p$ collisions at
$\sqrt{s}=8\, TeV$ and $\sqrt{s}=13\, TeV$ with $\lambda=-\,0.015$ 
($G_{PS}(0) = 0.00389\,GeV^{-1}$).\\
\end{center}
\begin{center}
\begin{tabular}{l c c }
\hline
channel & $\sigma\,fb,\,8\,TeV $ & $\sigma\,fb,\,13\,TeV$\\
\hline
$W^+\,X_{PS}$ & 24.1   & 59.7  \\
\hline
$W^-\,X_{PS}$ & 6.94 & 20.4 \\
\hline
$Z\,X_{PS}$ & 10.9 & 29.2 \\
\hline
$\gamma\,X_{PS}$ & 3.55 & 9.19 \\
\hline
$q(\bar q)\,jets\,X_{PS}$ & 152.3 & 338.8 \\
\hline
$W^+\,W^-\,X_{PS}$ & 27.3   & 749.3  \\
\hline
$W^+\,Z\,X_{PS}$ & 51.0 & 537.7 \\
\hline
$W^-\,Z\,X_{PS}$ & 12.3 & 138.2 \\
\hline
$W^+\,\gamma\,X_{PS}$ & 3.03 & 11.5 \\
\hline
$W^-\,\gamma\,X_{PS}$ & 0.78 & 2.86 \\
\hline
 
$\sigma(X_{PS})_{tot}$ & 292.3 & 1896.9 \\
\hline
\end{tabular}
\end{center}
\end{table}

Thus we consider possible pseudoscalar neutral resonance with mass $\approx 750\,TeV$, which mainly decays into
\begin{equation}
W^+W^-;\quad Z Z;\quad Z \gamma,\quad \gamma \gamma
;\label{eq:WWZ}
\end{equation}
and also to rare channels~(\ref{ll}). 
According to Table 1 the cross-section of the resonance production at
$\sqrt{s}=8\,TeV$ is six times less than at $\sqrt{s}=13\,TeV$. Available $\sqrt{s}=8\,TeV$ data~\cite{CMS13,PRL14,CMS15} do not contradict to our estimates with account of branching ratios~(\ref{WIDTH}).
Namely,
for $\sqrt{s} =8\,TeV$ we have
\begin{equation}
\sigma(p\,p \to X_{PS})\cdot BR(X_{PS} \to \gamma\,\gamma) =
292.3 \cdot 0.012\,=\,3.51\,fb;\label{gg8}
\end{equation}
that do not contradict the most recent limitations~\cite{PR15}. Limitations
for $W\,W$ and $Z\,Z$ decay modes~\cite{ATEPJ,AT815} also do not contradict the results.
For example,
CMS data~\cite{CMS15} give for $750\, GeV$ resonance with width $\simeq 100\,GeV$
limitation $\sigma BR(X_{PS}\to \gamma \gamma)< 40 \, fb$ with prediction ~(\ref{gg8}). Let us note, that
our result~(\ref{ll}) for channel $X_{PS} \to l^+ l^-l^+ l^-$
($l = e, \mu$)
with integral luminosity $L = 5.3\, fb^{-1}$~\cite{CMS13} gives the following estimate for the number of events
\begin{eqnarray}
& &\sigma(X_{PS},8\,TeV)\cdot BR(X_{PS}  \to l^+ l^-l^+ l^-)
\cdot L =\nonumber\\
& &292.3 \cdot 0.00065 \cdot 5.3 =  1.01\,.
\end{eqnarray}
It is worth mentioning, that in experimental results at $\sqrt{s}=8\,TeV$~\cite{CMS13} there is one event just at $M(l^+ l^-l^+ l^-) = 750\, GeV$ and no other events for $M(l^+ l^-l^+ l^-) > 600\, GeV$. Of course this coincidence proves nothing due to the poor
statistics, we may only state, that results~\cite{CMS13} do not contradict
our estimates.

Now what for $\sqrt{s} =13\,TeV$?
First  of all let us estimate an effect in channel $\gamma\,\gamma$.
We have for possible number of events with
(\ref{WIDTH}) and data from Table 1
\begin{equation}
N_{\gamma\gamma}\,=\,\sigma(p\,p \to X_{PS})\cdot BR(X_{PS} \to \gamma\,\gamma) \cdot L
 = 22.7\cdot L(fb^{-1}).\label{NGG}
\end{equation}
Thus we have for $L \simeq 3\,fb^{-1}$ few tens events, that agrees observations~\cite{750A,750C}.

It may be advisable to study effect not only in channel $X_{PS} \to \gamma\,\gamma$ but also in channel $X_{PS} \to \gamma\,l^+\,l^-$.
According to~(\ref{WIDTH}) we have
\begin{equation}
\frac{BR(X_{PS} \to \gamma\,l^+\,l^-)}{BR(X_{PS} \to \gamma\,\gamma)}\,=\,
\frac{0.67}{1.5}\,=\,0.45\,;\label{XGl}
\end{equation}
that is actually only two times smaller than already observed effect in $2\,\gamma$.

Let us also calculate the effect for $l^+\,l^-\,l^+\,l^-$ at $\sqrt{s} = 13\,GeV$ in the resonance region $\simeq 750\,GeV$
\begin{equation}
N(l^+\,l^-\,l^+\,l^-)\,=\,\sigma(X_{PS},13\,TeV)\cdot BR(X_{PS}  \to l^+ l^-l^+ l^-)\cdot L =
  1.23\cdot L(fb^{-1}).\label{4l}
\end{equation}
So even for $L \sim 10\,fb^{-1}$ the effect in the four leptons channel may become noticeable.
The more so as for this channel background conditions are favorable~\cite{CMS13}.
Effects $X_{PS} \to \gamma\,l^+\,l^-;\,X_{PS} \to l^+\,l^-\,l^+\,l^-$ with intensities~(\ref{XGl}, \ref{4l}) would
confirm definitely the interpretation of the $750\, GeV$ state being W-ball.
Note, that existing limitations on a possible extra contribution of decay
$X_{PS} \to \gamma Z$ with invisible decay
$Z \to \bar \nu \nu$~\cite{CMSgam1,ATLAS16} do not contradict our estimates.

Let us remind, that all the estimates were made with $\lambda = -0.015$. 
Calculations for another value of $\lambda $ are straightforward with 
prescriptions of the present work.

\section{Conclusion}

Existence of W-balls would testify for anomalous gauge interaction~(\ref{FFF}),
which would be due to non-perturbative effects in the electroweak interaction.
Thus we could come to important conclusion, that non-perturbative
contributions are appropriate not only to QCD, but to the electroweak interaction as well. In this case the anomalies in the electroweak boson pair production
do not contradict the Standard Model and do not need extra efforts for a
choice of a theory beyond the SM.

Data on effects under discussion might give information on a value of
parameter $\lambda$.
According to our considerations it could be expected in range $\lambda = -0.015\pm 0.005$. Of course experiments on direct measurement of $\lambda$,
{\it e.g.} in processes of $W^+\,W^-, W^\pm\, Z(\gamma)$  production are also quite desirable.

We would emphasize, that in case of a success, the wouldbe fact of a simultaneous matching of two difficult for explanation effects in the framework of our approach would be quite instructive. 
It might serve as a confirmation of non-perturbative
method~\cite{BAA09,AVZ2} in case of realization of the predictions being discussed here. This approach could serve for achieving of the additional information on links between fundamental parameters of the Standard Model (see {\it e.g.}~\cite{AVZ2}).

\section{Acknowledgments}

The work is supported in part by the Russian Ministry of Education and Science
under grant NSh-3042.2014.2.

\end{document}